\documentclass[10pt,twocolumn]{IEEEtran}
\usepackage{graphicx}
\usepackage{amsmath}
\usepackage{amssymb}
\usepackage[caption=false]{subfig}
\usepackage[noadjust]{cite}
\usepackage{float}
\usepackage{algorithm}
\usepackage{bibentry}
\usepackage{balance}

\newcounter{storeeqcounter}
\newcounter{tempeqcounter}

\begin{document}

\title{Low-Complexity Adaptive Beam and Channel Tracking for Mobile mmWave Communications}

\author{Yavuz Yap{\i}c{\i} and \.{I}smail G\"{u}ven\c{c}, \textit{Senior~Member, IEEE}\\
Department of Electrical and Computer Engineering, North Carolina State University, Raleigh, NC\\
Email: \{yyapici, iguvenc\}@ncsu.edu
\thanks{This work is supported in part by NASA grant NNX17AJ94A.}}%

\maketitle

\begin{abstract}
In this paper, we study low-complexity algorithms for beam and channel tracking for millimeter-wave (mmWave) communications. In particular, the least mean squares (LMS) and bidirectional LMS (BiLMS) algorithms are derived for a mobile mmWave transmission scenario, where channel measurement is a \textit{nonlinear} function of the unknown angle-of-arrival (AoA) and angle-of-departure (AoD). Numerical results confirm that LMS is superior to widely used Extended Kalman Filter (EKF) algorithm in tracking the mmWave beam, when the initialization of AoA/AoD and channel gains is imperfect (i.e., performed using noisy channel estimates). Moreover, BiLMS exhibits a very good mean square error (MSE) performance as compared to both LMS and EKF, which makes it a promising channel tracking algorithm for a mobile mmWave transmission scenario. We also show that LMS and BiLMS algorithms are more robust against the impairments due to the non-optimal antenna array size as compared to EKF, and show relatively faster convergence characteristic along with increasing signal-to-noise ratio (SNR). 
\end{abstract}


\section{Introduction}\label{sec:intro}
The millimeter-wave (mmWave) communications has attracted much attention in recent years as a promising candidate for future 5G wireless networks~\cite{Andrews14What5G,Heath14FiveDis,Rappaport2014MimMmW}. In order to combat severe path-loss in mmWave frequency bands, various beamforming techniques based on directional transmission have been investigated~\cite{Heath2017ExpSpaCha, Molisch2017HybBeaSur, Molisch2017OptChaSta}. Since efficient beamforming strategies require accurate channel state information (CSI) together with associated angle-of-arrival (AoA) and angle-of-departure (AoD), these unknowns should be continuously estimated as the channel varies. Assuming mobile communications with relatively fast variation, continuous estimation brings a huge computational complexity.

One common way to obtain a coarse estimate of the mmWave channel is to employ compressive sensing (CS) algorithms owing to the sparsity of these channels in both delay and angular domains~\cite{Heath2014ChaEst, Wang2018MilWav, Lee2016ChaEst}. The computational complexity and signaling overhead associated with the CS algorithms are, however, prohibitively high, and it is imperative to employ these algorithms frequently enough as the channel variation increases. The tracking algorithms has therefore received attention lately in the context of channel and/or beam tracking for mmWave communications.

In~\cite{Guo2016TraAng}, an analog beamforming strategy is adopted, where time-varying AoA/AoD and quasi-static path gain of a mmWave system are estimated using Extended Kalman Filter (EKF). A similar problem is considered in~\cite{Heath2016BeaTra} together with time-varying path gains, and the overall complexity of EKF based approach is further relieved. The beamforming optimization for a similar problem is discussed in~\cite{Kim2017RobBea}, where only one-dimensional angle estimation is performed via EKF assuming the equality of AoA and AoD. An interesting recent study of \cite{Love2018AdaBea} questions the optimality of the linear approximation that EKF makes while deriving the Jacobian matrix, and proposes Unscented Kalman Filter (UKF) as an extension, which might be a separate direction for this work, as well.     

In this paper, we consider tracking of unknown CSI and AoA/AoDs parameters for mmWave communications based on their previous values. In particular, least mean squares (LMS) algorithm is derived from the original steepest descent algorithm~\cite{Widrow85AdapSigProc}, where channel observations are nonlinear function of AoA and AoDs. Numerical results show that LMS has a better beam tracking capability than EKF algorithm whenever the initialization is imperfect, which is a realistic assumption considering channel acquisition algorithms~\cite{Yapici2018mmWpOMP}. In addition, a powerful extension of LMS, referred to as bidirectional LMS (BiLMS)~\cite{Yapici12BiLMS, Yapici2018TraPer}, is also shown to better track the channel as compared to EKF for both perfect and imperfect initialization.  

The rest of the paper is organized as follows. In Section~\ref{sec:system}, we introduce the system model together with the beamforming strategy. The LMS and BiLMS algorithms are derived in Section~\ref{sec:lms_tracking}, and the associated numerical results are presented in Section~\ref{sec:results}. The paper concludes with Section~\ref{sec:conclusion}. 

\vspace{-0.0in}

\section{System Model}\label{sec:system}
\subsection{Time-Varying mmWave Channel Model}
We assume a geometrical time-varying channel model for a mobile mmWave communication scenario given at time $k$ as
\begin{align} \label{eq:channel}
\textbf{H}_k = \sum\limits_{\ell=1}^{L} \alpha_{k,\ell} \textbf{a}_{\rm R}(\theta_{k,\ell}) \textbf{a}_{\rm T}(\phi_{k,\ell})^{\rm H} ,
\end{align} 
where $L$ is the number of multipaths, $\alpha_{k,\ell}$ is the complex path gain following the standard complex Gaussian distribution, $\theta_{k,\ell}$ and $\phi_{k,\ell}$ are AoA and AoD, respectively. Given $M$ and $N$ being the number of receive and transmit antennas, respectively, the array response vectors are given as follows
\begin{align}
\!\!\!\textbf{a}_{\rm R}(\varphi) &{=}\, \frac{1}{\sqrt{M}} \!\left[ 1 \; e^{{-}2j\pi \frac{d}{\lambda}\cos\left( \varphi \right) } \dots e^{{-}j2\pi \frac{d}{\lambda}\left( M{-}1\right)\cos\left( \varphi  \right) } \right]^{\rm T} \!\!\!\!, \\
\!\!\!\textbf{a}_{\rm T}(\varphi) &{=}\, \frac{1}{\sqrt{N}} \!\left[ 1 \; e^{{-}2j\pi \frac{d}{\lambda}\cos\left( \varphi \right) } \dots e^{{-}j2\pi \frac{d}{\lambda}\left( N{-}1\right)\cos\left( \varphi  \right) } \right]^{\rm T} \!\!\!\!,
\end{align}
where $d$ is the antenna spacing in the uniform linear array (ULA) and $\lambda$ is the wavelength.

The time evolution of the complex path gain is modeled by a first order auto-regressive (AR) process as follows~\cite{Heath2016BeaTra, Kim2017RobBea}
\begin{align}\label{eqn:time_evol_path}
\boldsymbol{\alpha}_{k{+}1} = \rho \boldsymbol{\alpha}_{k} + \textbf{u}^{\alpha}_k,   
\end{align} 
where $\boldsymbol{\alpha}_k \,{=}\, \left[ \alpha_{k,1} \, \alpha_{k,2} \dots \alpha_{k,L} \right]^{\rm T}$, $\rho$ is the correlation coefficient, and $\textbf{u}^{\alpha}_k$ is the respective innovation noise following a zero-mean complex Gaussian noise with covariance $\left(1{-}\rho^2\right) \textbf{I}_L$ with $\textbf{I}_L$ being the identity matrix of size $L{\times}L$. Furthermore, angular variation over time is modeled as a Gaussian noise process given as~\cite{Guo2016TraAng, Heath2016BeaTra, Kim2017RobBea}
\begin{align}
\boldsymbol{\theta}_{k{+}1} = \boldsymbol{\theta}_{k} + \textbf{u}^{\theta}_k, \label{eqn:time_evol_aoa}\\
\boldsymbol{\phi}_{k{+}1} = \boldsymbol{\phi}_{k} + \textbf{u}^{\phi}_k, \label{eqn:time_evol_aod}
\end{align} 
where $\boldsymbol{\theta}_k \,{=}\, \left[ \theta_{k,1} \, \theta_{k,2} \dots \theta_{k,L} \right]^{\rm T}$, $\boldsymbol{\phi}_k \,{=}\, \left[ \phi_{k,1} \, \phi_{k,2} \dots \phi_{k,L} \right]^{\rm T}$, and $\textbf{u}^{\theta}_k$ and $\textbf{u}^{\phi}_k$ follow independent zero-mean complex Gaussian processes with covariance $\sigma^2_{\theta}\textbf{I}_L$ and $\sigma^2_{\phi}\textbf{I}_L$, respectively.       

\subsection{Beamforming Model and Channel Observations}
\begin{figure}[!t]
	\centering
	\hspace*{-0.0in}
	\includegraphics[width=0.47\textwidth]{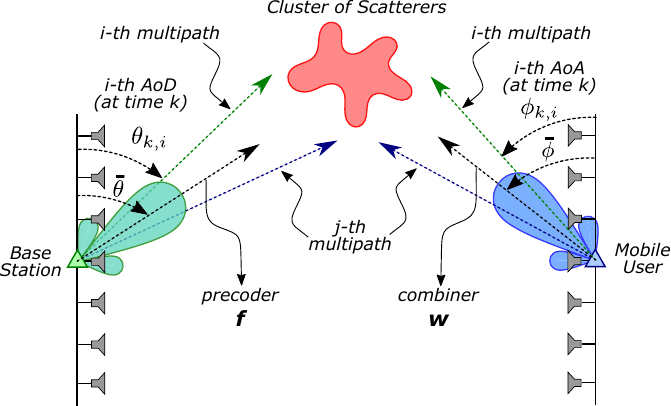}
	\label{fig:antenna_LMS}\vspace{-0.0in}
	\caption{Mobile mmWave communications setting with particular precoder and combiner vectors, and a single cluster of scatterers.}
	\label{fig:setting}
\end{figure}

Because of high energy consumption and cost of analog-to-digital converters (ADCs), we assume single radio-frequency (RF) chain at both end of the transceiver employing a single ADC. We therefore consider a fully analog transceiver with the transmit precoder $\textbf{f}$ and receive combiner $\textbf{w}$ as in Fig.~\ref{fig:setting}, which are given as
\begin{align}
\textbf{f} &\,{=}\, \frac{1}{\sqrt{N}} \!\left[ 1 \; e^{{-}2j\pi \frac{d}{\lambda}\cos\left( \overline{\phi} \right) } \dots e^{{-}j2\pi \frac{d}{\lambda}\left( N{-}1\right)\cos\left( \overline{\phi}  \right) } \right]^{\rm T} \!\!\!\!, \\
\textbf{w} &\,{=}\, \frac{1}{\sqrt{M}} \!\left[ 1 \; e^{{-}2j\pi \frac{d}{\lambda}\cos\left( \overline{\theta} \right) } \dots e^{{-}j2\pi \frac{d}{\lambda}\left( M{-}1\right)\cos\left( \overline{\theta}  \right) } \right]^{\rm T} \!\!\!\!,
\end{align}
where $\overline{\phi}$ and $\overline{\theta}$ are the precoder and combiner vector pointing directions, respectively. Note that beamforming with the precoder $\textbf{f}$ and combiner $\textbf{w}$ can be realized by progressively adjusting phase shifters~\cite{Heath2016BeaTra}. In addition, this formulation can be generalized to consider the multiple RF chains, as well, and possibly in a hybrid or full-digital transceiver structure instead, which will provide better control over pointing directions.

Assuming that $s_k$ is the unit-energy pilot symbol transmitted using the precoder $\textbf{f}$, and received symbols at each antenna are collected using the combiner $\textbf{w}$, observations are given as
\begin{align}\label{eq:observation}
y_k &= \textbf{w}^{\rm H} \textbf{H}_k \textbf{f} s_k + \textbf{w}^{\rm H} \textbf{v}_k,
\end{align}  
where $\textbf{v}_k$ is the noise vector following a zero-mean complex Gaussian process with covariance $N_0 \textbf{I}_M$. The associated signal-to-noise ratio (SNR) is given as $\mathsf{SNR} \,{=}\, M \! N/N_0$. Note that multiplying \eqref{eq:observation} by complex conjugate of $s_k$ eliminates the dependency on $s_k$ since $|s_k|^2\,{=}\,1$ while statistics of the modified noise $\tilde{v}_k\,{=}\,\textbf{w}^{\rm H} \textbf{v}_k$ remains the same. Hence, \eqref{eq:observation} can be equivalently written as
\begin{align}\label{eq:observation_equivalent_1}
y_k &= \sum\limits_{\ell=1}^{L} \alpha_{k,\ell} \textbf{w}^{\rm H} \textbf{a}_{\rm R}(\theta_{k,\ell}) \textbf{a}_{\rm T}(\phi_{k,\ell})^{\rm H} \textbf{f} + \tilde{v}_k .
\end{align}
Defining a function $g$ given as 
\begin{align}
g(K,\varphi) = \begin{cases}
\displaystyle 1 & \textrm{if } \varphi = 0, \\
\displaystyle \frac{1-e^{{-}2j\pi \frac{d}{\lambda}K\varphi}}{ K \left(  1-e^{{-}2j\pi \frac{d}{\lambda}\varphi} \right) } & \textrm{otherwise},
\end{cases}
\end{align}  
the observation model in \eqref{eq:observation_equivalent_1} becomes
\begin{align}
y_k &= \underbrace{\sum\limits_{\ell=1}^{L} \alpha_{k,\ell} \, g(N,\Delta\theta_{k\ell}) \,  g(M,\Delta\phi_{k\ell})}_{h \left( \textbf{x}_k \right)} + \tilde{v}_k , \label{eq:observation_equivalent_2}
\end{align}
where $\Delta\theta_{k\ell}\,{=}\,\cos(\theta_{k,\ell}){-}\cos(\overline{\theta})$, $\Delta\phi_{k\ell}\,{=}\,\cos(\overline{\phi}){-}\cos(\phi_{k,\ell})$, $h \left( \textbf{x}_k \right)$ is the measurement function with the unknown vector $\textbf{x}_k\,{=}\,[ \boldsymbol{\alpha}_{{\rm R},k}^{\rm T} \; \boldsymbol{\alpha}_{{\rm I},k}^{\rm T} \; \boldsymbol{\theta}_k^{\rm T} \; \boldsymbol{\phi}_k^{\rm T}]^{\rm T}$, where $\boldsymbol{\alpha}_{{\rm R},k}\,{=}\,{\rm Re}\{\boldsymbol{\alpha}_{k}\}$ and $\boldsymbol{\alpha}_{{\rm I},k}\,{=}\,{\rm Im}\{\boldsymbol{\alpha}_{k}\}$. In order to end up with real-valued equations only, we equivalently express \eqref{eq:observation_equivalent_2} as follows 
\begin{align}
\left[ {\begin{array}{c}
   y_{{\rm R},k} \\
   y_{{\rm I},k} \\
\end{array} } \right] &= \left[ {\begin{array}{c}
   h_{\rm R}(\textbf{x}_k) \\
   h_{\rm I}(\textbf{x}_k) \\
\end{array} } \right] + \left[ {\begin{array}{c}
   {\rm Re}\{\tilde{v}_{k}\} \\
   {\rm Im}\{\tilde{v}_{k}\} \\
\end{array} } \right], 
\end{align}   
where $y_{{\rm R},k}\,{=}\,{\rm Re}\{y_k\}$, $y_{{\rm I},k}\,{=}\,{\rm Im}\{y_k\}$, $h_{\rm R}(\textbf{x}_k)\,{=}\,{\rm Re}\{h(\textbf{x}_k)\}$, $h_{\rm I}(\textbf{x}_k)\,{=}\,{\rm Im}\{h(\textbf{x}_k)\}$, and $\textbf{h}(\textbf{x}_k)\,{=}\,[ h_{\rm R}(\textbf{x}_k) \; h_{\rm I}(\textbf{x}_k)]^{\rm T}$.

\section{Adaptive Beam and Channel Tracking}\label{sec:lms_tracking}
In this section, we present LMS and BiLMS algorithms to estimate unknown channel parameters (i.e., path gains, AoAs, AoDs) to track the mmWave channel in~\eqref{eq:channel} and/or the associated beam. 
\subsection{LMS Based Beam Tracking}
We first consider the LMS algorithm to adaptively track beam and channel for the communication scenario in Section~\ref{sec:system}. Since the measurement in \eqref{eq:observation_equivalent_2} is not linear in the unknown parameter vector $\textbf{x}_k$, we need to derive LMS using the original steepest descent algorithm given as
\begin{align}\label{eq:steepest_descent}
\hat{\textbf{x}}_{k+1} = \hat{\textbf{x}}_{k} - \boldsymbol{\mu} \boldsymbol{\nabla}_{\hat{\textbf{x}}_k} J_k ,
\end{align}
where $\hat{\textbf{x}}_{k}$ is the estimate of $\textbf{x}_{k}$, $\boldsymbol{\nabla}$ is the gradient operator, $J_k$ is the mean square error (MSE) at time instant $k$. We represent the adaptation step-size of the algorithm in \eqref{eq:steepest_descent} by the diagonal matrix $\boldsymbol{\mu} \,{=}\, {\rm diag} \big[ \mu_\alpha \textbf{1}_{2L} \; \mu_\theta \textbf{1}_{L} \; \mu_\phi \textbf{1}_{L} \big]$, where $\mu_\alpha$, $\mu_\theta$, and $\mu_\phi$ represent the particular step-size values of the channel path gain, AoA, and AoD, respectively. Defining $e_{{\rm R},k}$ and $e_{{\rm I},k}$ as the real and imaginary part of the estimation error $e_k\,{=}\,y_k\,{-}\,h(\hat{\textbf{x}}_k)$, the MSE is given as $J_k\,{=}\,\mathsf{E}\left\lbrace \| \textbf{e}_k \|^2 \right\rbrace$ with $\textbf{e}_k\,{=}\,[e_{{\rm R},k} \; e_{{\rm I},k}]^{\rm T}$. 

The gradient of the MSE is given as 
\begin{align}\label{eq:gradient_true}
\boldsymbol{\nabla}_{\hat{\textbf{x}}_k} J_k &= \frac{\partial }{\partial \hat{\textbf{x}}_k} \mathsf{E}\left\lbrace e_{{\rm R},k}^2 + e_{{\rm I},k}^2 \right\rbrace, \\
&= {-}2 \mathsf{E}\left\lbrace e_{{\rm R},k} \frac{\partial h_{\rm R}(\hat{\textbf{x}}_k)}{\partial \hat{\textbf{x}}_k} + e_{{\rm I},k} \frac{\partial h_{\rm I}(\hat{\textbf{x}}_k)}{\partial \hat{\textbf{x}}_k} \right\rbrace,\\
&= {-}2 \mathsf{E}\left\lbrace \textbf{e}_{k}^{\rm T} \frac{\partial \textbf{h}(\hat{\textbf{x}}_k)}{\partial \hat{\textbf{x}}_k} \right\rbrace.
\end{align} 

Note that the LMS algorithm originates from the steepest descent where the true gradient $\boldsymbol{\nabla}_{\hat{\textbf{x}}_k}$ is simply replaced by its instantaneous value $\hat{\boldsymbol{\nabla}}_{\hat{\textbf{x}}_k}$, which is given as~\cite{Widrow85AdapSigProc}
\begin{align}\label{eq:gradient_instantaneous}
\hat{\boldsymbol{\nabla}}_{\hat{\textbf{x}}_k} J_k = {-}2 \textbf{e}_{k}^{\rm T} \frac{\partial \textbf{h}(\hat{\textbf{x}}_k)}{\partial \hat{\textbf{x}}_k}, 
\end{align}
and the LMS algorithm is accordingly described as
\begin{align}\label{eq:lms}
\hat{\textbf{x}}_{k+1} = \hat{\textbf{x}}_{k} + 2 \boldsymbol{\mu} \textbf{e}_{k}^{\rm T} \frac{\partial \textbf{h}(\hat{\textbf{x}}_k)}{\partial \hat{\textbf{x}}_k}.
\end{align}
The derivative $\partial \textbf{h}(\hat{\textbf{x}}_k)/\partial \hat{\textbf{x}}_k$ in \eqref{eq:lms} is given as
\begin{align}\label{eq:derivative_h}
\!\!\!\!\frac{\partial \textbf{h}(\hat{\textbf{x}}_k)}{\partial \hat{\textbf{x}}_k} \,{=} \!\!
\left[  {\begin{array}{cccc}
\!\!\scalebox{1.2}{$\frac{\partial h_{\rm R}(\hat{\textbf{x}}_k)}{\partial \boldsymbol{\alpha}_{{\rm R},k}}$} &\!\! \scalebox{1.2}{$\frac{\partial h_{\rm R}(\hat{\textbf{x}}_k)}{\partial \boldsymbol{\alpha}_{{\rm I},k}}$} &\!\! \scalebox{1.2}{$\frac{\partial h_{\rm R}(\hat{\textbf{x}}_k)}{\partial \boldsymbol{\theta}_k}$} &\!\! \scalebox{1.2}{$\frac{\partial h_{\rm R}(\hat{\textbf{x}}_k)}{\partial \boldsymbol{\phi}_k}$}\!\!\\[8pt]
\!\!\scalebox{1.2}{$\frac{\partial h_{\rm I}(\hat{\textbf{x}}_k)}{\partial \boldsymbol{\alpha}_{{\rm R},k}}$} &\!\! \scalebox{1.2}{$\frac{\partial h_{\rm I}(\hat{\textbf{x}}_k)}{\partial \boldsymbol{\alpha}_{{\rm I},k}}$} &\!\! \scalebox{1.2}{$\frac{\partial h_{\rm I}(\hat{\textbf{x}}_k)}{\partial \boldsymbol{\theta}_k}$} &\!\! \scalebox{1.2}{$\frac{\partial h_{\rm I}(\hat{\textbf{x}}_k)}{\partial \boldsymbol{\phi}_k}$}\!\!
\end{array} } \right],
\end{align}  
where
\begin{align}
\frac{\partial h(\hat{\textbf{x}}_k)}{\partial \alpha_{{\rm R},k\ell}} &= g(N,\Delta\theta_{k\ell}) \,  g(M,\Delta\phi_{k\ell}) \label{eq:derivative_hR}\\
\frac{\partial h(\hat{\textbf{x}}_k)}{\partial \alpha_{{\rm I},k\ell}} &= j g(N,\Delta\theta_{k\ell}) \,  g(M,\Delta\phi_{k\ell}) \label{eq:derivative_hI} \\
\frac{\partial h(\hat{\textbf{x}}_k)}{\partial \theta_{k,\ell}} &= \alpha_{k,\ell} \frac{\partial g(N,\Delta\theta_{k\ell})}{\partial \theta_{k,\ell}} \,  g(M,\Delta\phi_{k\ell}) \label{eq:derivative_aoa}\\
\frac{\partial h(\hat{\textbf{x}}_k)}{\partial \theta_{k,\ell}} &= \alpha_{k,\ell} g(N,\Delta\theta_{k\ell}) \,  \frac{\partial g(M,\Delta\phi_{k\ell})}{\partial \phi_{k,\ell}} \label{eq:derivative_aod}
\end{align}
with the general derivative of function $g$ given in \eqref{eq:derivative_angle} (on the top of next page) for $\Delta\varphi\,{\neq}\,0$, and being $0$ otherwise. Note that the angle derivatives in \eqref{eq:derivative_angle} can be shown to be $\partial \Delta\theta_{k\ell}/\partial \theta_{k,\ell}\,{=}\,{-}\sin\left(\theta_{k,\ell}\right)$ and $\partial \Delta\phi_{k\ell}/\partial \phi_{k,\ell}\,{=}\,\sin\left(\phi_{k,\ell}\right)$ for \eqref{eq:derivative_aoa} and \eqref{eq:derivative_aod}, respectively. Note also that first two columns of \eqref{eq:derivative_h} can be obtained readily by taking real and imaginary parts of \eqref{eq:derivative_hR} and \eqref{eq:derivative_hI}. 

\setcounter{storeeqcounter}{\value{equation}}

\begin{figure*}[!h]
	\normalsize
	\setcounter{tempeqcounter}{\value{equation}} 
	\begin{align}\label{eq:derivative_angle}
	\setcounter{equation}{\value{storeeqcounter}} 
	\frac{\partial g(K,\Delta\varphi)}{\partial \varphi} = j \frac{2\pi\frac{d}{K\lambda}\frac{\partial \Delta\varphi}{\partial \varphi}}{\left(1-e^{{-}2j\pi \frac{d}{\lambda}\Delta\varphi }\right)^2 } \left[ K e^{{-}2j\pi \frac{d}{\lambda} K\Delta\varphi} - e^{{-}2j\pi \frac{d}{\lambda}\Delta\varphi} - (K{-}1) e^{{-}2j\pi \frac{d}{\lambda}(K{+}1)\Delta\varphi} \right] 
	\end{align}
	\setcounter{equation}{\value{tempeqcounter}} 
	\hrulefill
	\vspace*{0pt}
\end{figure*}

\addtocounter{equation}{1}

\subsection{BiLMS Based Channel Tracking}
Each new beam alignment and channel estimation can be considered as an additional initialization at the end of the tracking period provided that no abrupt change occurs in channel features. This approach enables the use of the BiLMS algorithm which is given as
\begin{align}
\hat{\textbf{x}}_{k+1}^{\rm F} &= \hat{\textbf{x}}_{k}^{\rm F} + 2 \boldsymbol{\mu} \textbf{e}_{k}^{\rm T} \frac{\partial \textbf{h}(\hat{\textbf{x}}_k^{\rm F})}{\partial \hat{\textbf{x}}_k^{\rm F}} ,\label{eq:bilms_f}\\
\hat{\textbf{x}}_{k-1}^{\rm B} &= \hat{\textbf{x}}_{k}^{\rm B} + 2 \boldsymbol{\mu} \textbf{e}_{k}^{\rm T} \frac{\partial \textbf{h}(\hat{\textbf{x}}_k^{\rm B})}{\partial \hat{\textbf{x}}_k^{\rm B}} ,\label{eq:bilms_b}
\end{align}
where $\hat{\textbf{x}}_{k}^{\rm F}$ and $\hat{\textbf{x}}_{k}^{\rm B}$ are the forward and backward estimates of unknown parameter vector $\textbf{x}_{k}$, and the final estimate is 
\begin{align}
\hat{\textbf{x}}_{k} = \frac{\left(\hat{\textbf{x}}_{k}^{\rm F} + \hat{\textbf{x}}_{k}^{\rm B} \right)}{2}. \label{eq:bilms_final}
\end{align} 
Note that the bidirectional processing in BiLMS needs an initial value of the unknown channel (i.e., channel path gains together with AoA/AoD values) at the end of the tracking period to initialize the backward adaptations in \eqref{eq:bilms_b}. As a result, the BiLMS algorithm is not suitable for beam tracking while a decision loop is conducted to evaluate whether any update in beamforming directions are required or not. However, once a new full channel estimation is performed and new AoA/AoD values are identified, the BiLMS algorithm can be used to improve the estimates within the most recent tracking period, which will be shown in the next section to be a promising approach.

\section{Numerical Results}\label{sec:results}
In this section, we investigate the performance of the LMS and BiLMS algorithms together with the EKF algorithm~\cite{Guo2016TraAng, Heath2016BeaTra, Kim2017RobBea} in tracking the mmWave beam and/or channel. We present results for AoA only, since AoD is statistically equivalent to AoA. We assume a multi-antenna system with $M\,{=}\,16$, $N\,{=}\,16$, and SNR${=}30$ dB unless otherwise stated. The time evolution of the underlying mmWave channel is governed by $\rho\,{=}\,0.995$ and $\sigma_\theta^2\,{=}\,\sigma_\phi^2\,{=}\,(0.5^\circ)^2$ in \eqref{eqn:time_evol_path}-\eqref{eqn:time_evol_aod}, which corresponds to a fast angular variation~\cite{Heath2016BeaTra}. We assume a single multipath with $L\,{=}\,1$, which is very likely for a narrow mmWave physical beam, and that the precoder and combiner vectors are pointing an arbitrary direction of $45^\circ$ (i.e., $\overline{\theta}\,{=}\,\overline{\phi}\,{=}\,45^\circ$). In addition, initial value of AoA (AoD) is assumed to take a continuous value uniformly within $10^\circ$ angular spread around the combiner (precoder) direction (i.e., $\theta_{1,\ell} \,{\in}\, \mathcal{U}\left[\,\overline{\theta}{-}5^\circ,\overline{\theta}{+}5^\circ \right]$\footnote{$\mathcal{U}\left[a,b\right]$ denotes real-valued uniform distribution in [$a,b$] with $a,b\,\in \mathbb{R}$.}). We use the best step-size value for the LMS and BiLMS adaptations based on trial and error, which have been found to be $\mu_\alpha \,{=}\, 0.1$ and $\mu_\theta \,{=}\, \mu_\phi \,{=}\, 0.0001$ for the path gain and AoA/AoD, respectively. 

\begin{figure}[!h]
	\centering
	\hspace*{-0.0in}
	\subfloat[Perfect Initialization]{\includegraphics[width=0.5\textwidth]{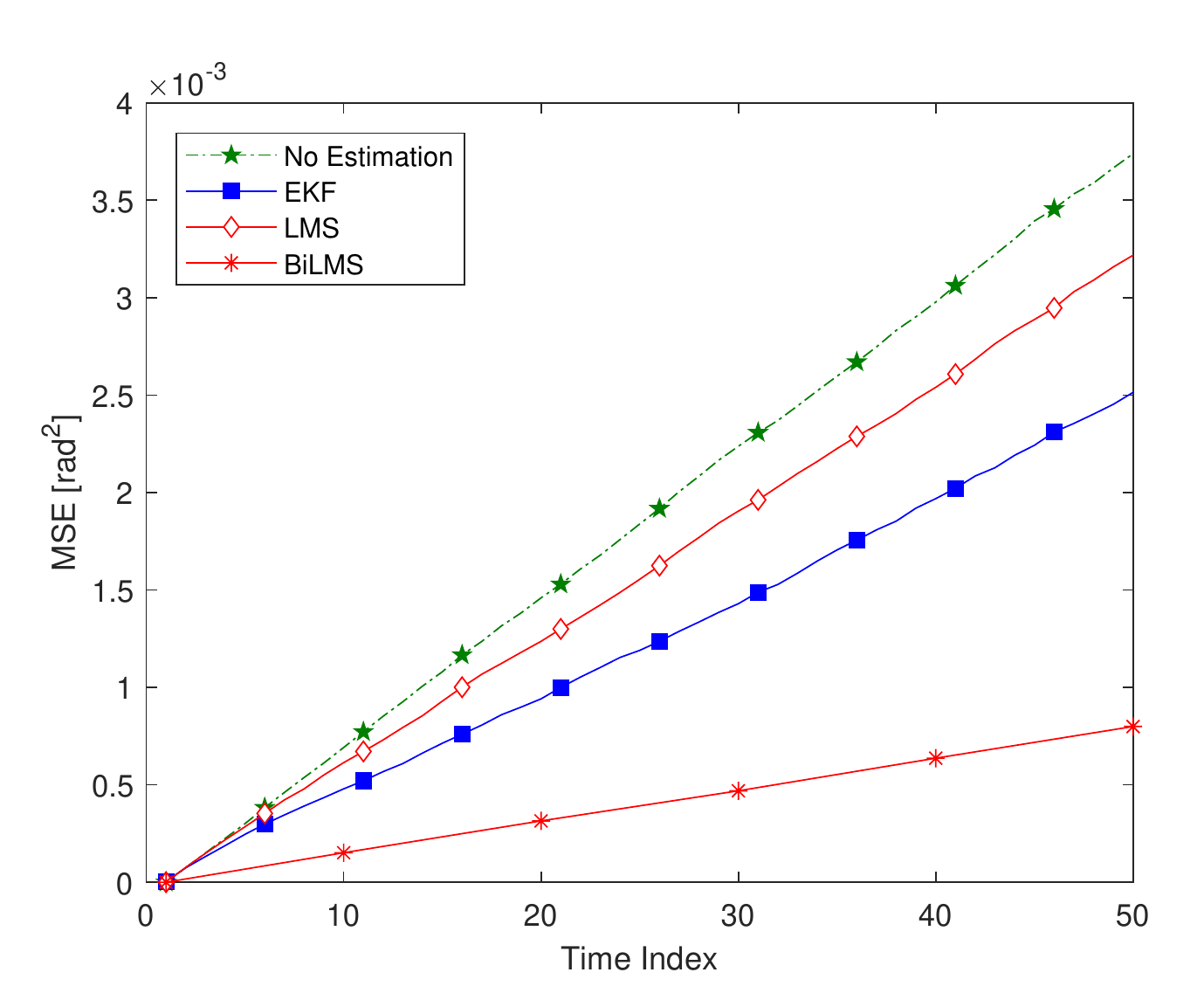}
		\label{fig:initialization_perfect}}\vspace{-0.25in}\\
	\subfloat[Imperfect Initialization]{\includegraphics[width=0.5\textwidth]{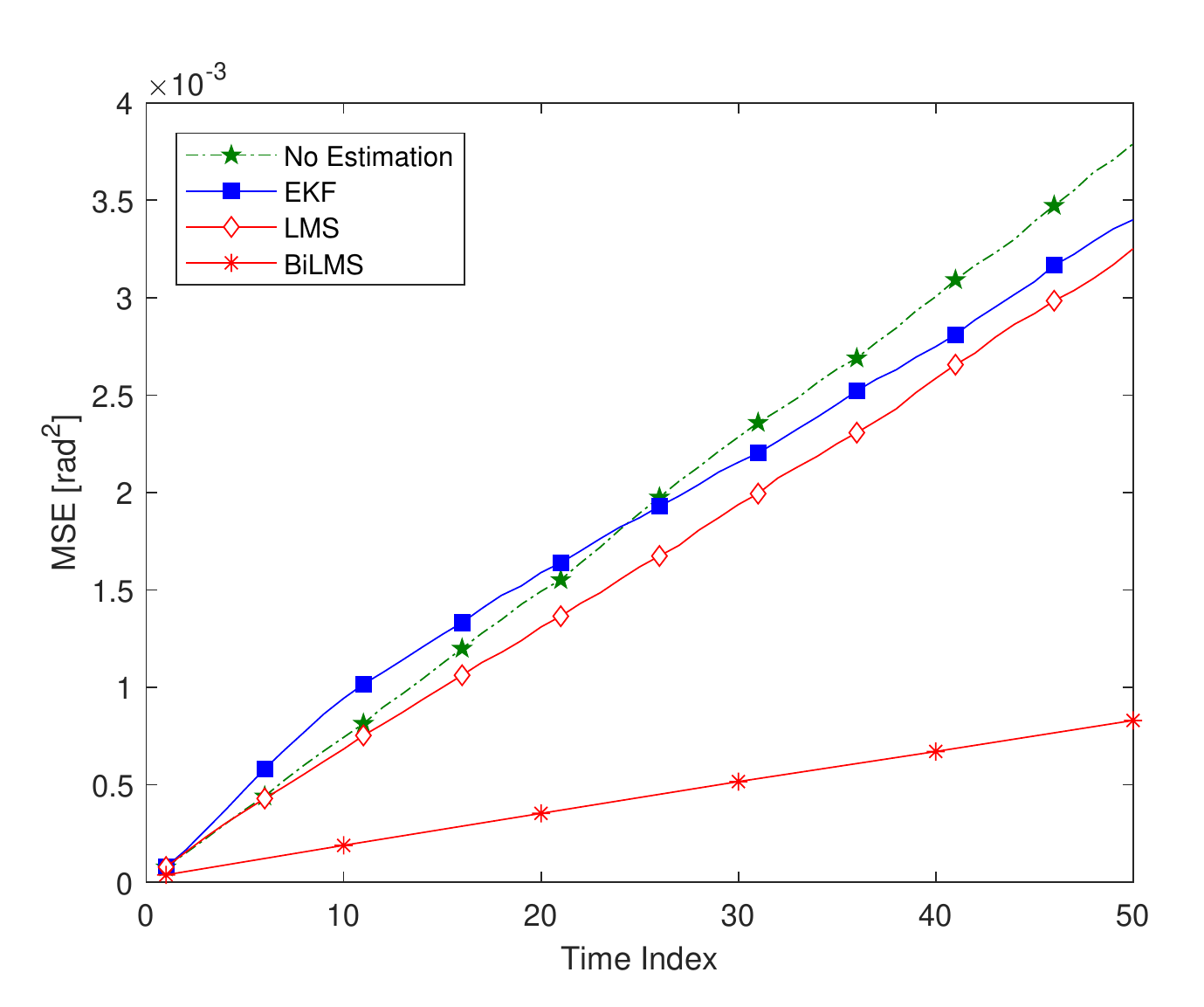}
		\label{fig:initialization_imperfect}}\vspace{-0.0in}
	\caption{MSE of LMS, BiLMS, and EKF for perfect and imperfect initialization together with no estimation performance, $16{\times}16$ antennas, and SNR$\,{=}\,30$ dB.}
	\label{fig:initialization}
\end{figure}

In Fig.~\ref{fig:initialization}, the evolution of average MSE for AoA is depicted along with discrete time index for the LMS, BiLMS, and EKF algorithms. In particular, we consider perfect (i.e., $\hat{\textbf{x}}_1\,{=}\,\textbf{x}_1$) and imperfect (i.e., $\hat{\textbf{x}}_1\,{=}\,\textbf{x}_1 + \boldsymbol{\epsilon}$) initialization schemes in Fig.~\ref{fig:initialization}\subref{fig:initialization_perfect} and Fig.~\ref{fig:initialization}\subref{fig:initialization_imperfect}, respectively. We assume that initialization error is represented by a real-valued random vector $\boldsymbol{\epsilon}$, which is zero-mean Gaussian with the covariance $\boldsymbol{\Lambda} \,{=}\, {\rm diag} \big[ \sigma^2_{\epsilon,\alpha} \textbf{1}_{2L} \; \sigma^2_{\epsilon,\psi} \textbf{1}_{2L} \big]$, where $\textbf{1}_{2L}$ is a length-$2L$ all-ones vector. For this particular setting, we choose path gain variance $\sigma^2_{\epsilon,\alpha}$ and angle variance $\sigma^2_{\epsilon,\psi}$ (due to imperfect initialization) to be $(0.5)^2$ and $(0.5^\circ)^2$, respectively. In addition, we also include the result for the case where no estimation algorithm is employed at all (i.e., $\hat{\textbf{x}}_k\,{=}\,\hat{\textbf{x}}_1$ for $\forall k$) as a practical benchmark. 

We observe in Fig.~\ref{fig:initialization}\subref{fig:initialization_perfect} that the tracking performance of EKF outperforms LMS when the initialization is perfect, which is a pretty unrealistic assumption. In addition, the MSE performance of BiLMS is superior to that of EKF when AoA is allowed to be estimated with further processing along the recent tracking period. On the other hand, Fig.~\ref{fig:initialization}\subref{fig:initialization_imperfect} shows that the performance of EKF significantly degrades for more realistic imperfect initialization assumption, and becomes worse than that of either LMS or BiLMS. We also observe that both LMS and BiLMS are very robust against initialization imperfections by showing only a slight performance degradation in comparison to perfect initialization scenario. 

\begin{figure}[!h]
	\centering
	\hspace*{-0.0in}
	\subfloat[LMS and BiLMS]{\includegraphics[width=0.5\textwidth]{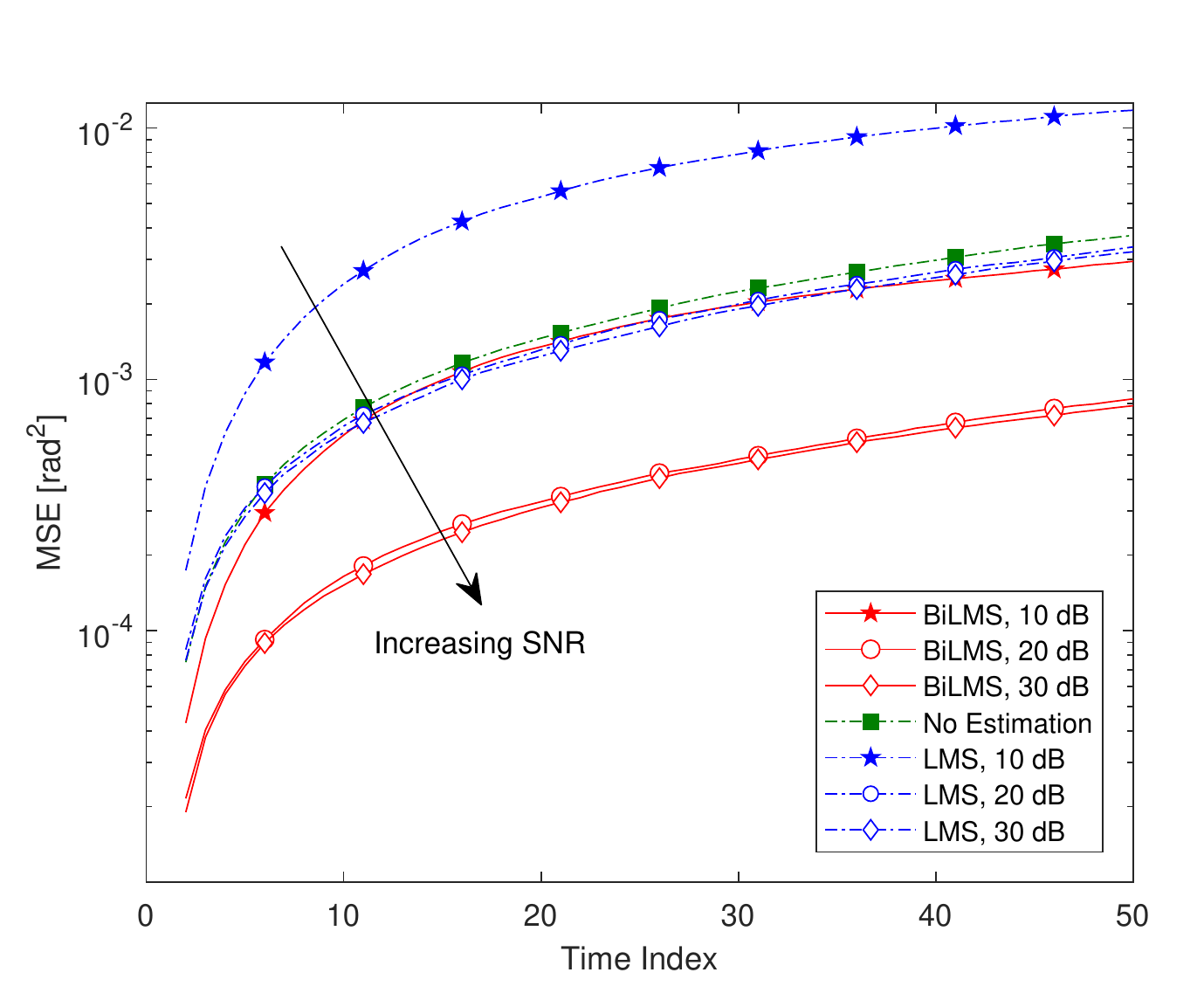}
		\label{fig:snr_LMS}}\vspace{-0.25in}\\
	\subfloat[EKF]{\includegraphics[width=0.5\textwidth]{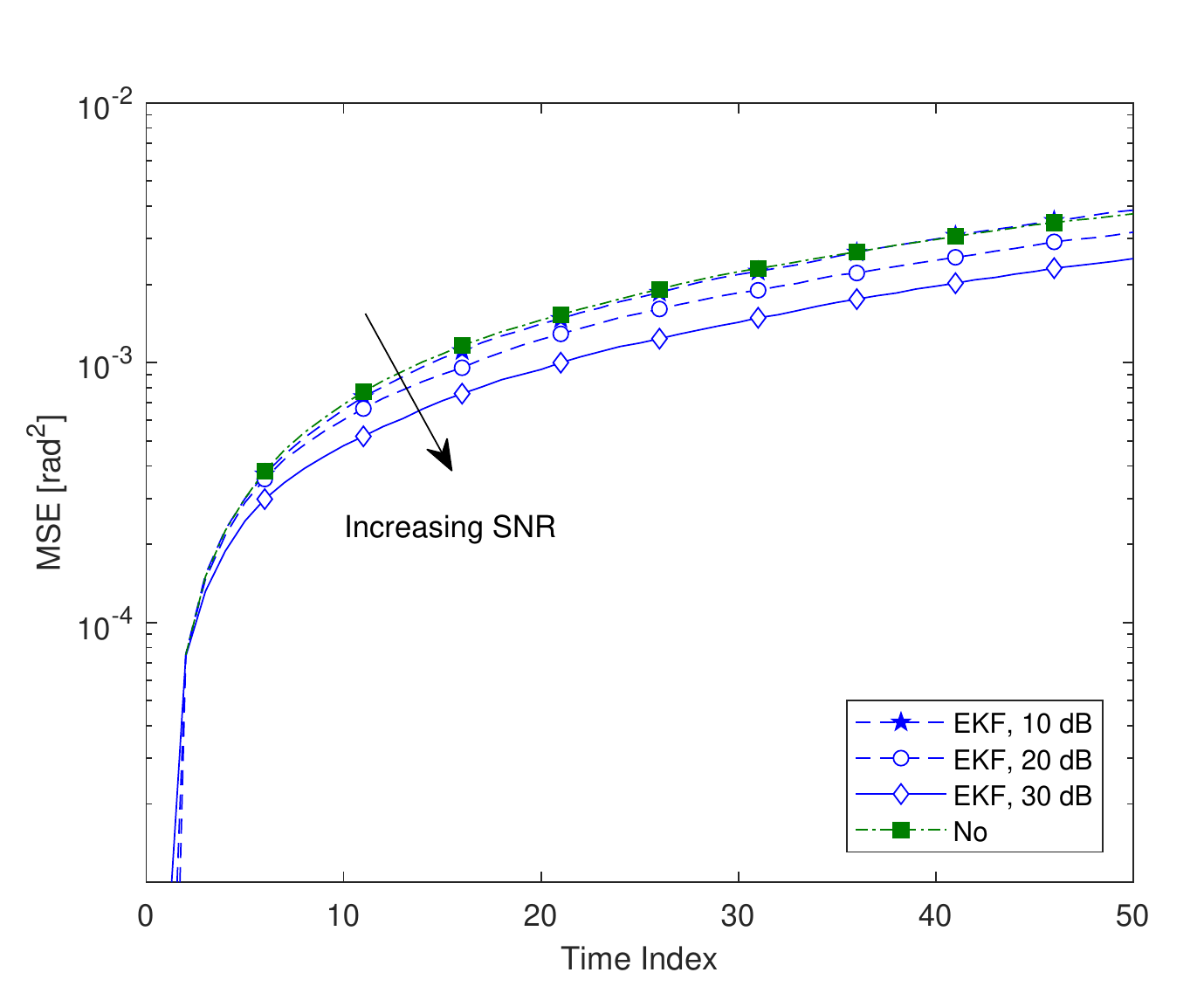}
		\label{fig:snr_EKF}}\vspace{-0.0in}
	\caption{MSE performance of LMS, BiLMS, and EKF along with varying SNR assuming perfect initialization and $16{\times}16$ antennas.}
	\label{fig:snr}
\end{figure}

In Fig.~\ref{fig:snr}, we depict the MSE performance of the algorithms under consideration for varying SNR values and assuming perfect initialization. We observe in Fig.~\ref{fig:snr}\subref{fig:snr_LMS} that although LMS and BiLMS both perform poorly when SNR is insufficient (i.e., SNR$\,{=}\,10\,\text{dB}$), they rapidly improve when SNR reaches up to $20\,\text{dB}$. Any performance improvement associated with even higher SNR values (i.e., SNR$\,{=}\,30\,\text{dB}$) is observed to remain marginal. On the other hand, the performance of EKF keeps improving as SNR continues to increase towards $30\,\text{dB}$. Although LMS and BiLMS pretty much saturate at SNR$\,{=}\,20\,\text{dB}$, the performance gap of EKF between $20\,\text{dB}$ and $30\,\text{dB}$ of SNR is still at a nonignorable level.  

\begin{figure}[!h]
	\centering
	\hspace*{-0.0in}
	\subfloat[LMS and BiLMS]{\includegraphics[width=0.5\textwidth]{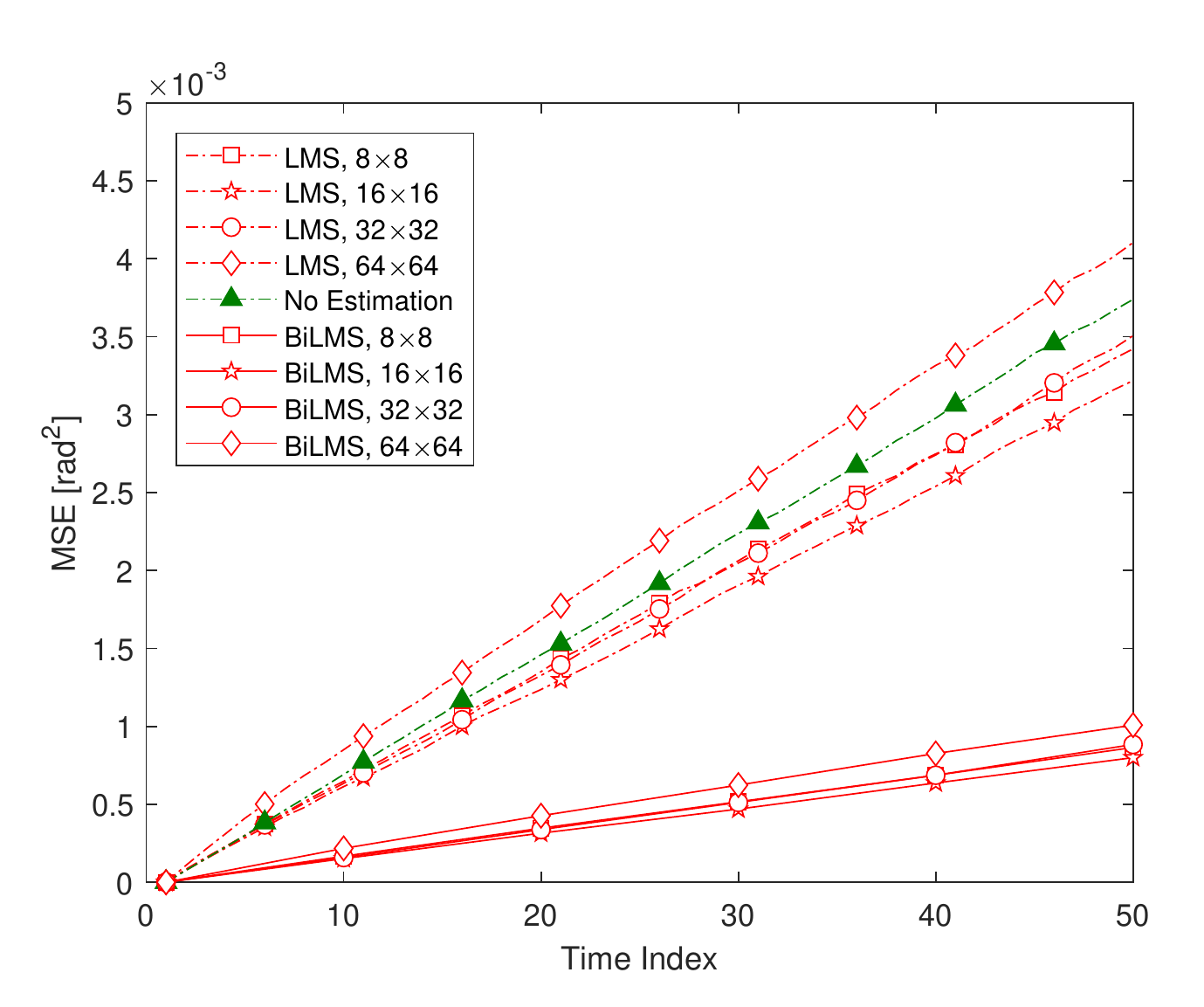}
		\label{fig:antenna_LMS}}\vspace{-0.25in}\\
	\subfloat[EKF]{\includegraphics[width=0.5\textwidth]{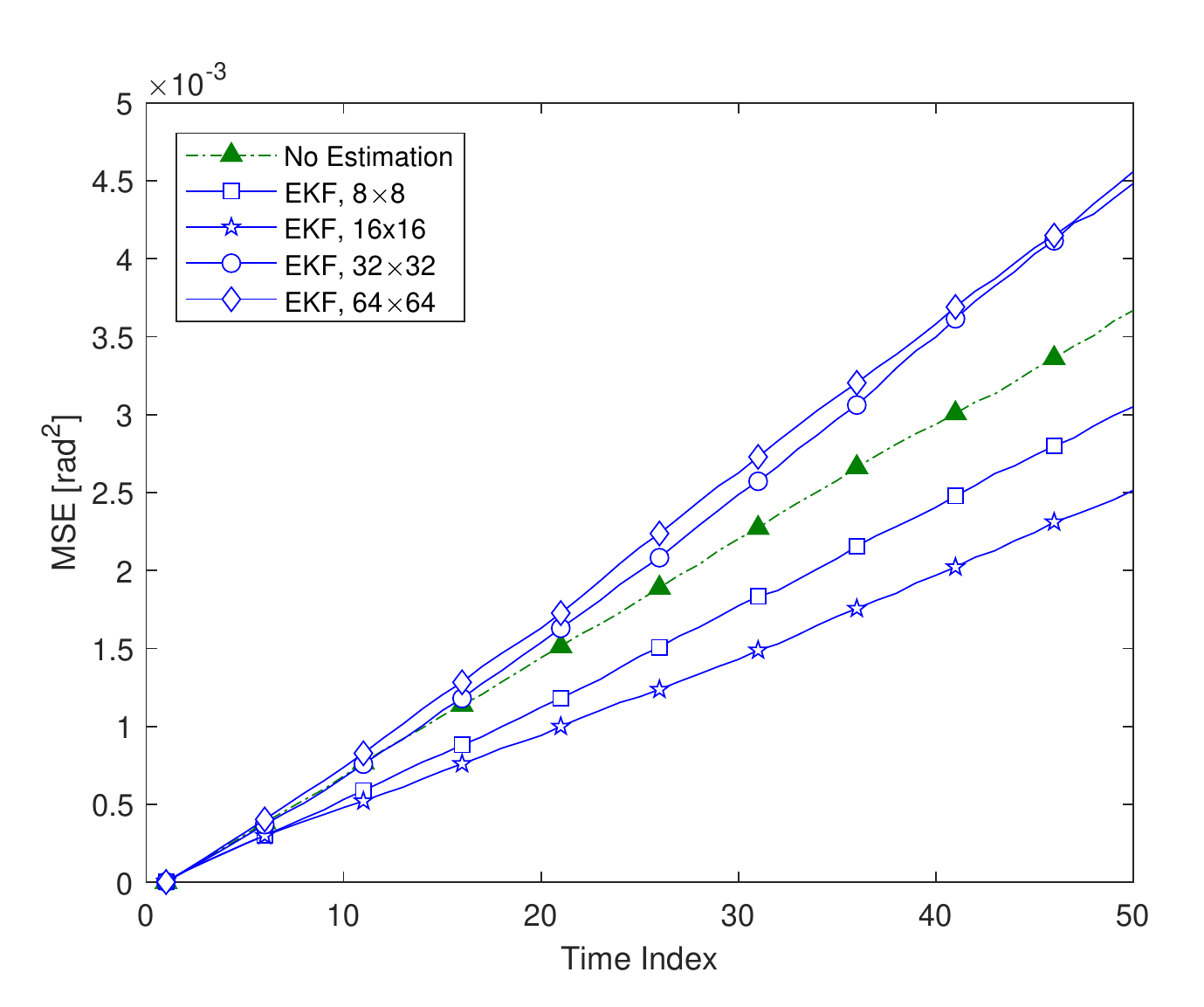}
		\label{fig:antenna_EKF}}\vspace{-0.0in}
	\caption{MSE performance of LMS, BiLMS, and EKF along with varying number of antennas assuming perfect initialization and SNR$\,{=}\,30$ dB.}
	\label{fig:antenna}
\end{figure}

Finally, Fig.~\ref{fig:antenna} demonstrates the effect of antenna array size on the MSE performance of the algorithms under consideration. We observe that the array size of $16$ at both transmit and receive sides achieves the best performance. This result underscores the compromise between small and large antenna array sizes~\cite{Heath2016BeaTra} (i.e., small arrays are favorable in tracking applications with their wide beam while they lose power with their small array factor gain, and vice versa). We also notice that the performance of EKF degrades severely as a result of not optimal array size choice, while the respective performance degradation for LMS and BiLMS seem to be relatively smaller.    

\balance 

\section{Conclusion}\label{sec:conclusion}
We consider tracking of channel and physical beam in a mobile mmWave communications scenario. We derived the LMS algorithm and its extension BiLMS from the original steepest descent algorithm under a nonlinear observation model. Numerical results show that LMS and BiLMS are very robust to imperfections (i.e., initialization with noisy channel estimates, employing non-optimal antenna array size), and rapidly converge to their best performance along with increasing SNR. On the other hand, we observe that the EKF algorithm is very vulnerable to aforementioned imperfections, and exhibit relatively slow convergence along with increasing SNR.

\bibliographystyle{IEEEtran}
\bibliography{IEEEabrv,paperbib}

\end{document}